\documentclass[12pt]{iopart}
\usepackage{iopams}
\usepackage[dvips]{graphicx}
\usepackage{epsfig}
\usepackage{tabularx}
\usepackage{pifont}
\usepackage{color}
\usepackage[pdfpagemode=UseNone,colorlinks=true,linkcolor=blue,citecolor=blue]{hyperref}

\newcommand{\nm}{\bar{n}}
\newcommand{\nth}{\bar{n}_{\mathrm{th}}}
\newcommand{\nbac}{\bar{n}_{\mathrm{BA}}^{cool}}
\newcommand{\nbap}{\bar{n}_{\mathrm{BA}}^{probe}}
\newcommand{\Geff}{\Gamma_{\mathrm{eff}}}
\newcommand{\Gopt}{\Gamma_{\mathrm{opt}}}
\newcommand{\area}{\mathcal{A}}
\newcommand{\Temp}{T_{bath}}

\newcommand{\meff}{m_{\mathrm{eff}}}
\newcommand{\pct}{\%}

\begin{document}

\title{Calibrated quantum thermometry in cavity optomechanics}

\author{A Chowdhury$^{1,2}$, P Vezio$^{2,3}$, M Bonaldi$^{4,5}$, A Borrielli$^{4,5}$, F Marino$^{1,2}$, B Morana$^{4,6}$, G Pandraud$^{6}$, A Pontin$^{7}$, G A Prodi$^{5,8}$, P M Sarro$^{6}$, E Serra$^{5,6}$ and F Marin$^{1,2,3,9}$}
\address{$^1$ CNR-INO, L.go Enrico Fermi 6, I-50125 Firenze, Italy}
\address{$^2$ Istituto Nazionale di Fisica Nucleare (INFN), Sezione di Firenze, Via Sansone 1, I-50019 Sesto Fiorentino (FI), Italy}
\address{$^3$ European Laboratory for Non-Linear Spectroscopy (LENS), Via Carrara 1, I-50019 Sesto Fiorentino (FI), Italy}
\address{$^4$ Institute of Materials for Electronics and Magnetism, Nanoscience-Trento-FBK Division,
 38123 Povo, Trento, Italy}
\address{$^5$ Istituto Nazionale di Fisica Nucleare (INFN), Trento Institute for Fundamental Physics and Application, I-38123 Povo, Trento, Italy}
\address{$^6$ Dept. of Microelectronics and Computer Engineering /ECTM-EKL, Delft University of Technology, Feldmanweg 17, 2628 CT  Delft, The Netherlands}
\address{$^7$ Department of Physics and Astronomy, University College London, Gower Street, London WC1E 6BT, United Kingdom}
\address{$^8$ Dipartimento di Fisica, Universit\`a di Trento, I-38123 Povo, Trento, Italy}
\address{$^{9}$ Dipartimento di Fisica e Astronomia, Universit\`a di Firenze, Via Sansone 1, I-50019 Sesto Fiorentino (FI), Italy}
\ead{marin@fi.infn.it}

\begin{abstract}

Cavity optomechanics has achieved the major breakthrough of the preparation and observation of macroscopic mechanical oscillators in peculiarly quantum states. The development of reliable indicators of the oscillator properties in these conditions is important also for applications to quantum technologies. We compare two procedures to infer the oscillator occupation number, minimizing the necessity of system calibrations. The former starts from homodyne spectra, the latter is based on the measurement of the motional sidebands asymmetry in heterodyne spectra. Moreover, we describe and discuss a method to control the cavity detuning, that is a crucial parameter for the accuracy of the latter, intrinsically superior procedure. 

\end{abstract}

\vspace{2pc}

\maketitle

\section{Introduction}

A crucial outcome of cavity optomechanics \cite{AKMreview} is the observation of peculiar quantum features in the behavior of macroscopic mechanical oscillators. 
The most relevant indicator of the achieved mechanical quantum domain is 
the so-called motional sidebands asymmetry. 
The optomechanical interaction generates spectral peaks around the carrier frequency of a probe field, at distances equal to the mechanical oscillation frequency $\Omega_m$. Their amplitudes are generally different according to quantum theory.
Different interpretations have been proposed to explain such asymmetry \cite{Khalili2012,Weinstein2014,Borkje2016}, all agreeing in recognizing it as a non-classical signature of the mechanical oscillator \cite{Borkje2016}, as soon as spurious experimental features are avoided \cite{Jayich2012,Safavi2013}. A particularly elucidating explanation considered that the anti-Stokes (blue) sideband implies an energy transfer from the oscillator to the field (frequency up-conversion of photons), and vice versa for the Stokes (red) sideband. Since the quantum oscillator cannot yield energy when it is in the ground state, the anti-Stokes process is less favored. It turns out that the blue and red sideband strengths are proportional respectively to $\nm$ and $(\nm+1)$ \cite{Wineland1987}, where $\nm$ is the mean occupation number of the oscillator.

Measurements of the sidebands asymmetry have been extensively used to monitor the motion of trapped ions \cite{Diedrich1989}, and it has recently become a key technique for cavity optomechanics. Besides its utility as direct indicator of the oscillator quantum behavior, the sidebands asymmetry is a powerful index to deduce the oscillator occupation number avoiding delicate evaluations of optomechanical parameters, such as oscillator effective mass or optomechanical gain, and calibrations of the detection system. It has been remarked that the thermal occupation number $\nth$ allows a direct evaluation of the absolute temperature, and it is therefore of extraordinary potential metrological interest. Several experiments concerning the use of optomechanical quantum effects for the measurement of absolute temperature, covering the full range from ultra-cryogenic to room temperatures, are indeed been running \cite{Purdy2017,Sudhir2017}. As a matter of fact, we expect that accurate measurements of the oscillator displacement variance and of motional sidebands asymmetry will be extensively exploited in the next future, and deserve a detailed investigation.

Sidebands asymmetry has been measured in optical experiments by alternatively positioning a probe field at a detuning of  
$\pm \Omega_m$ around the cavity resonance \cite{Safavi2012}, as well as, in a single measurement, from the spectral sidebands in a probe field \cite{Purdy2015,Underwood2015,Peterson2016,Sudhir2017a,Rossi2018}. The former technique is particularly useful in the regime of deeply resolved sidebands ($\Omega_m \gg \kappa$, where $\kappa$ is the cavity linewidth), since for each position of the probe the measured sideband is at the cavity resonance frequency and it is thus amplified. On the other hand, the control of systematic effects can be an issue: the system should remain stable between two separate measurement sessions, the probe intensity and the detection efficiency must be equal for the two values of detuning, and the probe detuning itself must be very accurate. The latter technique, while introduced more recently in cavity optomechanics, is already well established, but it also requires an accurate control of the probe detuning, above all in the case of narrow cavity resonance. The cavity works indeed as frequency filter for the output field, with an effect that differs between the two sidebands and can thus spoil the measurement of their ratio.

In this work we experimentally investigate the sidebands asymmetry as signature of quantum performance, and we compare it with a further indicator, i.e., the oscillator displacement variance measured form the area of the corresponding peak in the probe phase spectrum. Furthermore, we demonstrate a method for correcting the measured sidebands asymmetry for non-null probe detuning, exploiting the spectral features of the device oscillating modes that are weakly coupled to the cavity field (``heavy'' modes).       

\section{Theoretical background}

The displacement spectrum of a mechanical oscillator is characterized by resonance peaks corresponding to the different normal modes. The area underlying each peak is a measure of the variance of the motion of the harmonic oscillator associated to the readout of that normal mode. It can be written as $\mathcal{A}_x = 2 x_{\mathrm{ZPF}}^2 (1/2 + \nm)$ where $x_{\mathrm{ZPF}}=\sqrt{\hbar/2 m_{\mathrm{eff}} \Omega_m}$ is the zero-point fluctuations amplitude, and $\nm$ is the mean occupation number ($\Omega_m$ is the oscillator angular frequency, $m_{\mathrm{eff}}$ its the effective mass). If the oscillator is in thermal equilibrium with a background at temperature $\Temp$, the mean thermal occupation number is $\nth \simeq k_{B}\Temp / \hbar \Omega_m$ ($k_{B}$ is the Boltzmann constant, and this expression of $\nth$ is valid in the high temperature limit $\nth >> 1$), and the peak width is $\Gamma_m = \Omega_m/Q$, where $Q$ is the intrinsic mechanical quality factor.

When the mechanical oscillator is embedded in an optical cavity, the optomechanical interaction with intracavity radiation yields thermalization toward the photon bath at negligible occupation number (``back-action cooling'' \cite{Arcizet2006,Gigan2006}), at a rate $\Gopt$ proportional to the cooling laser power. The width of the spectral peak becomes $\Geff = \Gamma_m + \Gopt$ and the oscillator occupation number is reduced by a factor of $\Geff/\Gamma_m$. However, the back-action of the optomechanical measurement introduces an additional fluctuating force acting on the oscillator, that can be seen as the effect of the quantum noise in the radiation pressure. Since such quantum fluctuations are proportional to the laser power, and actually to $\Gopt$, the originated displacement noise of the optically damped oscillator has negligible dependence on the cooling power, in the limit $\Gopt >> \Gamma_m$. Its contribution to the total displacement variance can be written in terms of additional occupation number $\nbac$ as \cite{AKMreview,Marquardt2007}  
\begin{equation}
\nbac=\left[\frac{\mathcal{L}(\Delta_{cool}+\Omega_{m})}{\mathcal{L}(\Delta_{cool}-\Omega_{m})}-1\right]^{-1}    
\label{nbac}
\end{equation}
where $\mathcal{L}(\omega)=1/\left[(\kappa/2)^{2}+\omega^{2}\right]$ is the Lorentzian response function of the optical cavity with linewidth $\kappa$, and $\Delta_{cool}$ is the detuning of the cooling radiation with respect to the cavity resonance.

The oscillator motion implies variations of the optical cavity resonance frequency $\omega_{\mathrm{cav}}$, at the rate $G = -\partial \omega_{\mathrm{cav}}/\partial x$. Such frequency fluctuations can be measured by exploiting the optical field leaving the cavity. The readout of the oscillator motion may be performed by analyzing the same radiation used for cooling. However, such field is commonly strongly detuned from the cavity resonance to assure an efficient cooling, therefore the optical susceptibility of the cavity is not trivial to be accurately considered. It is more practical to introduce an additional, resonant probe field. The drawback is its additional back-action, that increases the oscillator noise. The probe back-action force does not depend on the cooling power, and it has the same effect of an increased background temperature. In general, the quantum radiation pressure noise produced by an intracavity field at detuning $\Delta$ is proportional to $\bar{n}_{\mathrm{cav}}^{max}\,\mathcal{L}(\Delta)\,\left[\mathcal{L}(\Delta+\Omega_m)+\mathcal{L}(\Delta-\Omega_m)\right]$ where $\bar{n}_{\mathrm{cav}}^{max}$ is the average number of intracavity photons in case of resonant field, that is proportional to the input power. This expression allows us to write the oscillator occupation number added by the probe beam in the form
\begin{equation}
\nbap=\nbac \frac{P^{probe}}{P^{cool}}\,\frac{\mathcal{L}(\Delta_{probe})}{\mathcal{L}(\Delta_{cool})}\,\frac{\mathcal{L}(\Delta_{probe}+\Omega_m)+\mathcal{L}(\Delta_{probe}-\Omega_m)}{\mathcal{L}(\Delta_{cool}+\Omega_m)+\mathcal{L}(\Delta_{cool}-\Omega_m)} 
\label{nbap}   
\end{equation}
where $P^{probe/cool}$ are the input powers of the probe/cool beam. Expressions (\ref{nbac}) and (\ref{nbap}) are particularly useful in the analysis of the experimental results, since they do not require the evaluation of the laser coupling efficiency and the consequent intracavity photon number, that are often not trivial. We remark that $\nbap$ is proportional to $1/P^{cool}$ and actually to $1/\Gopt$, provided that the probe beam is close to resonance and has therefore a negligible effect on the effective width.

In conclusion, the total effective occupation number can be written as 
\begin{equation}
\nm = \nth \frac{\Gamma_m}{\Geff}+\nbac+\nbap  \, .
\label{ntot}   
\end{equation}

A useful parameter to be considered is the area$\times$width product $\area \Geff$ of 
the spectral peak. In the classical limit, when the variance of the motion is still 
dominated by thermal noise, such product should remain constant as the 
cooling power is increased, keeping, in the displacement spectrum, the value of $\area_x \times \Geff \simeq 2 x_{\mathrm{ZPF}}^2 \nth \Gamma_m = k_B \Temp/\meff \Omega_m Q$. Quantum noise is instead at the origin of a linear increase of $\area \Geff$ 
versus $\Geff$. The peak area$\times$width product in the frequency spectrum can be written as 
\begin{equation}
\area \Geff = 2 g_0^2 \Gamma_m \left[\nth + \left(\nbac+\nbap+\frac{1}{2} \right)\frac{\Geff}{\Gamma_m}  \right]
\label{aw}
\end{equation}
where the vacuum optomechanical coupling strength is $g_0 = G x_{\mathrm{ZPF}}$.
The accurate independent estimate of $g_0$ is not straightforward, since it crucially relies on the readout calibration and laser coupling efficiency. On the other hand, the terms into square brackets in Eq. (\ref{aw}), i.e., operatively, the ratio between the slope and the intercept in the $\area \Geff$ vs $\Geff$ line, is directly linked to meaningful properties of the oscillator quantum state. It can be used as a check of the agreement between the expected and the measured behavior of the optomechanical system, i.e., to verify the absence of unmodeled extra noise, as well as, e.g., for evaluating $\nth$ and actually the oscillator thermodynamic temperature $\Temp$.

A more accurate measure of the oscillator occupation number can be obtained from the heterodyne spectra of the field reflected by the cavity, that allow to distinguish the two sidebands produced by the Stokes and anti-Stokes processes in the optomechanical interaction. For a resonant probe, the sidebands peaks have areas proportional respectively to $\nm$ (anti-Stokes) and $\nm + 1$ (Stokes), therefore $\nm$ is directly calculated from the Stokes to anti-Stokes sidebands ratio $R$ as $\nm = 1/(R-1)$. This indicator has at least two interesting properties: it does not require a calibration of the measured spectra in terms of, e.g., oscillator displacement or frequency fluctuations, and it is robust against effects of possible misleading extra-noise. It should be remark, however, that correlated phase and amplitude noise in the probe field can produce spurious sidebands asymmetry even at high occupation numbers \cite{Jayich2012}.  

A crucial concern for the sidebands thermometry is the residual detuning of the probe with respect to the cavity resonance. The two motional sidebands are indeed filtered by the cavity according to $\mathcal{L} (\Delta_{probe} \pm \Omega_m)$, and such filtering effect modifies $R$ as soon as $\Delta_{probe} \ne 0$, thus spoiling the validity of the measurement. The main original contribution of our work is indeed a method to control and evaluate such residual probe detuning and the consequent correction to the measured $R$.

\section{Experimental Setup}

The measurements are performed on a circular SiN membrane with a thickness
of $100\,$nm and a diameter of $1.64\,$mm, supported by a silicon
ring frame. This frame is suspended on four points
with alternating flexural and torsional springs, forming an on-chip ``loss shield'' structure \cite{Borrielli2014}. More information
about the design, fabrication and the characteristic of the device can be
found in Borrielli {\it et al.} \cite{Borrielli2016} and Serra {\it et al.} \cite{Serra2016,Serra2018}. 
The theoretical resonance frequencies of the drum modes in a circular membrane are given by the expression
$f_{mn} = f_0\,\alpha_{mn}$
where $\alpha_{mn}$ is the $n$-th root of the Bessel polynomial $J_m$ of order $m$, and
$
f_{0} = \frac{1}{\pi}\, \sqrt{\frac{T}{\rho}}\frac{1}{\Phi}
\label{f0}
$
($T$ is the stress, $\rho$ the density, $\Phi$ the diameter of the membrane). The measured frequencies are in close agreement (to better than $0.1 \pct$) with the theoretical expression, where at cryogenic temperature $f_0 = 96.6\,$kHz. For $m > 0$ we expect couples of degenerate modes. In the real device the perfect circular symmetry is broken, two orthogonal axes are defined and the two quasi-degenerate modes (the measured frequency split is below 100 Hz) have shapes nominally given by the expression $J_m (\alpha_{mn} r) \cos m \theta$ and $J_m (\alpha_{mn} r) \sin m \theta$, where $(r, \theta)$ are polar coordinates and $r$ is normalized to the membrane radius. 

The oscillator is
placed in a Fabry-Perot cavity of length $4.38$~mm, at $2$~mm from the cavity flat end mirror, forming a ``membrane-in-the-middle'' setup \cite{Jayich2008}. The input coupler is concave with a radius of 50~mm, originating a waist of 70~$\mu$m. The cavity finesse and linewidth are respectively $24500$ and $\kappa=1.4 \ \textrm{MHz} \times 2\pi$.
The cavity optical axis is displaced from the center of the membrane by $\sim 0.2$ mm, roughly along the axis with $\theta \simeq 0$. As a consequence, the optomechanical coupling and readout are much more efficient for one of the modes in each quasi-degenerate couple (with the shape $\propto \cos m \theta$), that we identify as ``light twin'', with respect to the other one (``heavy twin'').

In this work we mainly focus on the (1,1) modes at 370 kHz, having a quality factor of $8.9\times10^{6}$
at cryogenic temperature, which leads to an intrinsic width $\Gamma_{m}/2 \pi = 40\,$mHz.

\begin{figure}
\begin{centering}
\includegraphics[scale=0.5]{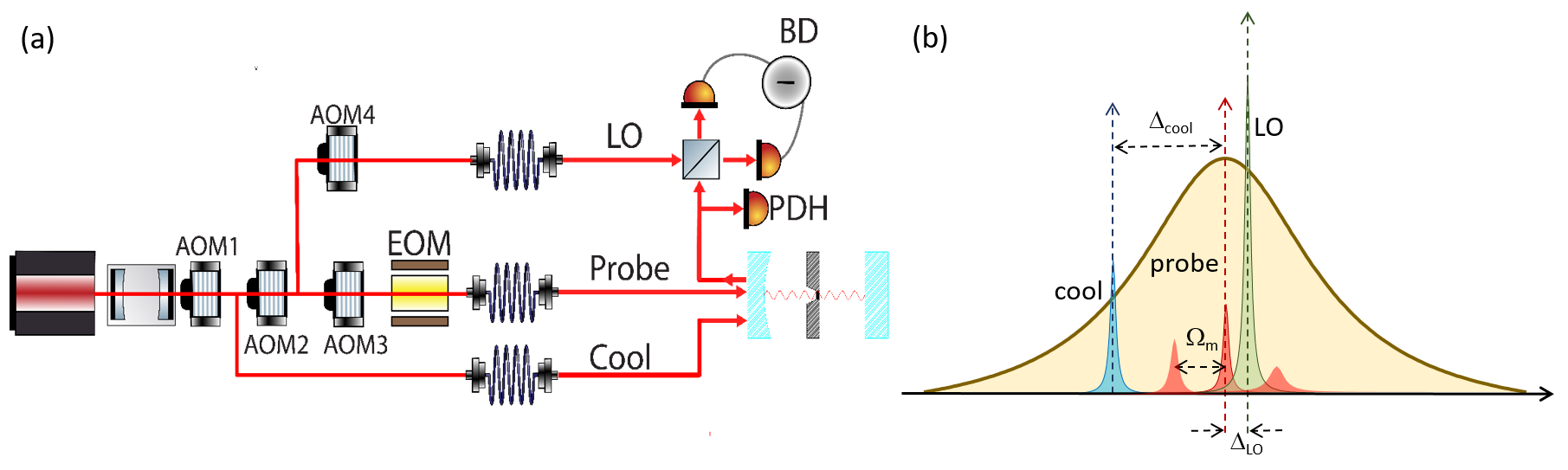}
\caption{(a) Simplified scheme of the experimental setup. The light of a Nd:YAG laser is filtered by a cavity having a linewidth of $66\,$kHz, frequency tuned by a first acousto-optic modulator (AOM), 
and split into three parts.
The first beam (probe) is frequency shifted by two cascade AOMs acting on opposite orders, and phase modulated by an electro-optic modulator (EOM) at $13\,$MHz for the 
Pounder-Drever-Hall (PDH) locking to the resonance of the optomechanical
cavity (OMC). The difference between the frequencies of the cascade AOMs defines the detuning of the second beam (cooling beam). The third beam (local oscillator, LO) is picked up after the second AOM, and frequency shifted by a fourth AOM. Its detuning with respect to the probe is defined by the frequency difference between the third and fourth AOMs.
After single-mode fibers, the first two beams are combined with orthogonal polarizations and mode-matched to the OMC. About $2 \mu$W of the reflected probe are sent to the PDH detection, while most of the probe light ($18 \mu$W typically impinge on the cavity) is combined with the LO in a balanced detection (BH). (b) Scheme of the beam frequencies. The LO is placed on the blue side of
the probe and detuned by $\Delta_{\mathrm{LO}} << \Omega_m$, therefore the Stokes lines are on the red
side of the LO, while the anti-Stokes lines are on the blue side. In the heterodyne spectra, they are located respectively at $\Omega_m+\Delta_{\mathrm{LO}}$ (Stokes) and $\Omega_m-\Delta_{\mathrm{LO}}$ (anti-Stokes).}
\label{setup}
\end{centering} 
\end{figure}

The optomechanical cavity is cooled down to $\sim 7\,$K in an helium flux cryostat. The light of a Nd:YAG laser is filtered by a Fabry-Perot cavity and split into three beams, whose frequencies are controlled by means of acousto-optic modulators (AOM) (Fig. \ref{setup}a).
The first
beam (probe) is always resonant with the optomechanical cavity, to which it is kept locked using the Pound-Drever-Hall detection and a servo loop. This exploits the first AOM to follow fast fluctuations, and a piezo-electric transducer to compensate for long term drifts of the cavity length. The second beam (cooling beam),
orthogonally polarized with respect to the probe, is also sent to the cavity and
red detuned by roughly half linewidth $(\Delta_{cool}= -2 \pi \times 700 \mathrm{kHz} \simeq -\kappa/2)$. The third beam is used as local oscillator (LO) in a balanced detection of the probe beam, reflected by the cavity. In such a detection
scheme the LO can either be frequency shifted with respect to the probe
(by $\Delta_{\mathrm{LO}}/2\pi \sim 9\,$kHz), allowing a low-frequency heterodyne detection \cite{Pontin2018a} (see Fig. \ref{setup}b), or phase-locked to the probe
for an homodyne detection of its phase quadrature. The first scheme (heterodyne) is useful to separate the motional sidebands  generated around the optical frequency of the probe field, at frequency shifts corresponding to the mechanical
modes frequencies. The spectra acquired with the second scheme (homodyne) are calibrated in terms of cavity frequency fluctuations using a calibration tone in the probe field, and are used to measure the variance of the motion of the different membrane normal modes.     

\section{Experimental results}

\begin{figure}
\begin{centering}
\includegraphics[scale=0.4]{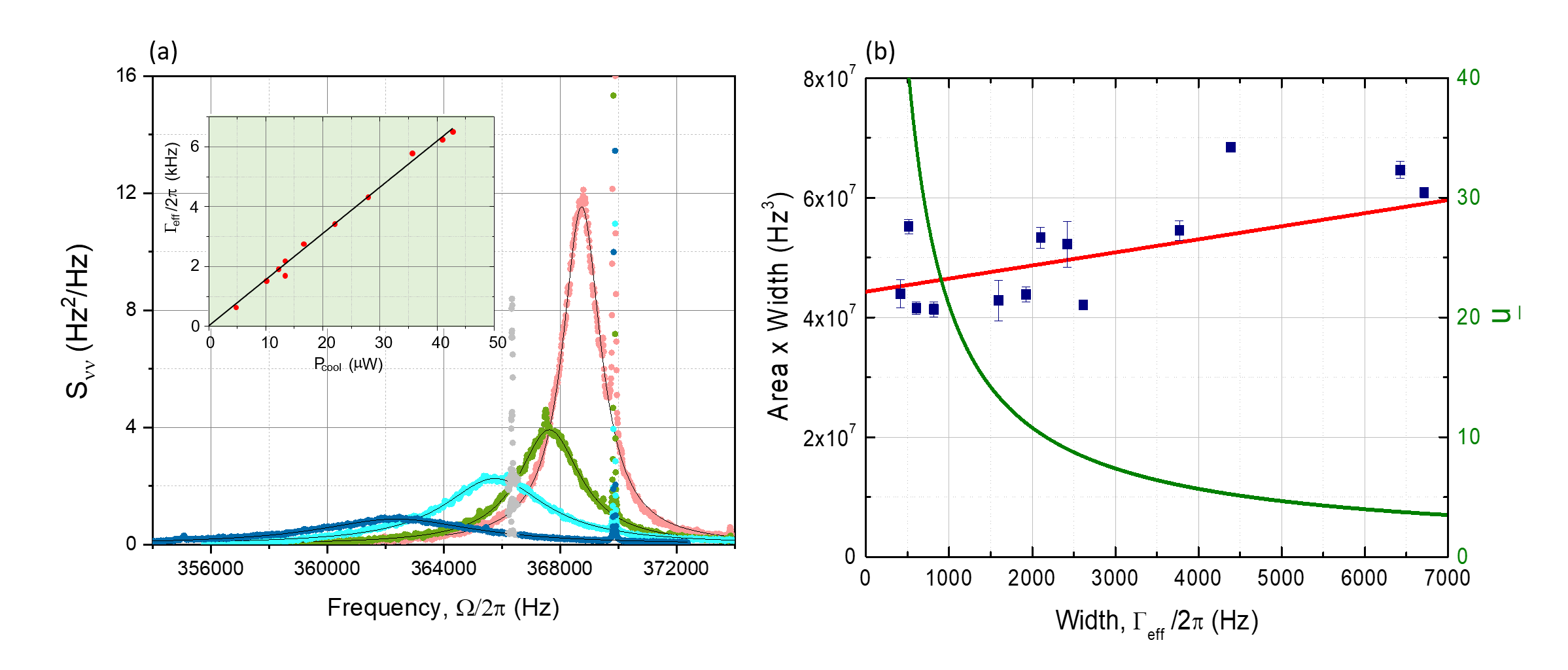}
\par\end{centering}
\caption{(a) Calibrated homodyne spectra around the frequency the (1,1) mechanical modes as the
cooling power is increased up to $\sim60 \mu$W, maintaining a detuning of $\Delta_{cool}\sim\kappa/2$. The peak associated to the ``light twin'' mode (strongly coupled to the radiation) exhibits the expected red shift (negative optical spring) and optical cold damping  (increase in its width $\Geff$ and decrease in the peak area). At $\sim370$ kHz is visible the narrow peak due to the ``heavy twin'' mode. Symbols show the experimental data, solid lines are the Lorentzian functions fitting the peaks of the ``light twin'' modes. A spurious electronic peak is shown with light grey symbols. The inset shows the measured peak width $\Geff/2 \pi$ as a function of the cooling power, together with a linear fit.
(b) Increment of the measured area$\times$width product for the strongly coupled (1,1) mode, as a function of its width $\Geff/2\pi$. 
The red straight line reports the prediction of Eq. (\ref{aw}), where just an overall scaling factor is fitted to the data. 
A solid green line shows the mean occupation number $\nm$ calculated according to Eq. (\ref{ntot}).}
\label{homo}
\end{figure}

We will focus on the (1,1) membrane modes around 370 kHz, and we will start our analysis from 
the homodyne spectra of our optomechanical system. The power of the cooling beam is increased by steps up to $\sim 60 \mu$W. The result,
as shown in Fig. \ref{homo}a, is a gradual cooling of the light mode
with a characteristic increase of $\Geff$ and a simultaneous red-shift of the
mechanical resonance frequency due to so-called
optical spring effect. On the other hand, the ``heavy twin'' mode is weakly coupled to the radiation since the optical spot is close to its nodal axis, therefore the associated spectral peak at 370 kHz shows negligible optomechanical effects.

The decrease of the peak area is an indication
of the reduction of the phonon occupancy $\nm$ in the ``light twin'' mode. A quantitative evaluation of $\nm$ from this parameter would require an independent measurement of the optomechanical gain. On the other hand, the cooling factor $\Geff/\Gamma_m$ can be accurately measured, but it provides a good estimate of the oscillator effective temperature and consequently of its occupation number $\nm \approx \nth \Gamma_m/\Geff$ just in the classical limit (as soon as the back-action is negligible) and in the absence of extra noise. The two indicators can be usefully put together in the area$\times$width product, that is shown in Fig. \ref{homo}b as a function of $\Geff$. The reported values of $\area \Geff$ are obtained from fits of the spectral peaks with a Lorentzian function. Eq. (\ref{aw}) predicts that the $\area \Geff$ vs $\Geff$ data should display a linearly increasing behavior, where the slope-to-offset ratio is determined by the different contributions to $\nm$. We have calculated such contributions using independent measurements, as follows. $\nth$ is calculated from the bath temperature measured by a silicon diode sensor fixed on the cavity, and the oscillator frequency. $\nbac$ is calculated from Eq. (\ref{nbac}) using the measured cavity linewidth and the detuning $\Delta_{cool}$ fixed with the AOMs. $\nbac$ is calculated from Eq. (\ref{nbac}) assuming $\Delta_{probe}\simeq 0$, measuring the probe-to-cooling beam power ratio and fitting the linear dependence between $P^{cool}$ and $\Geff$ (see the inset in Fig. \ref{homo}a). Finally, $\Gamma_m$ is obtained from ring-down measurements with an additional laser at 970 nm, where the measured optomechanical effects are very weak due to the low cavity finesse and laser power. The experimental measurements well follow the predicted slope, shown with a solid line in Fig. \ref{homo}b, where just the overall vertical scaling factor is fitted to the data. Here the error bars just reflects the standard deviation of measurements performed on consecutive acquisitions. The scattering of the data shows that longer term fluctuations in system parameters (when changing the cooling power) dominate over such statistical uncertainties, that are therefore not considered as meaningful in the following analysis.

A further solid curve in the figure shows the behavior of $\nm$ calculated from Eq. (\ref{ntot}), i.e., assuming that the system is well modeled and in the absence of additional noise, as suggested by the good agreement between the prediction of Eq. (\ref{aw}) and the experimental data. We infer that an occupation number of $\nm = 3.9$ is achieved at our maximum cooling power.
Moreover, the fitted vertical scaling factor allows to estimate, according to Eq. (\ref{aw}), a vacuum optomechanical coupling strength of $g_0/2\pi = 31 \pm 1$ Hz. We remark again that such additional inferred parameter is not involved in the evaluation of $\nm$.

The observed qualitative agreement is not yet a safe guarantee of an accurate measurement. Heating of the membrane oscillator due to laser absorption would yield a linear increase of $\Temp$ with $P_{cool}$, and thus a larger slope of $\area \Geff$ vs $\Geff$. Leaving the slope as free parameter in the fit of $\area \Geff$ vs $\Geff$, we find for the ratio between slope and offset a value of $(8.0 \pm 2.5) \times 10^{-5}$ Hz$^{-1}$, to be compared with $4.9 \times 10^{-5}$ Hz$^{-1}$ calculated from Eq. (\ref{aw}). This suggests that the sample temperature could have increased by $1.8 \pm 1.5$ K at our maximum cooling power. Furthermore, additional amplitude or frequency noise in the laser field would instead provide a quadratic term in $\area \Geff$ vs $\Geff$. 
We have indeed added such term to the fit of our data, finding a maximum contribution of  
$13 \pm 10\%$ to the measured $\area$. In both fitting procedures the uncertainty is due to the scattering of the experimental data, and the results are compatible with null effects of heating and extra laser noise.  

While the described analysis of the homodyne spectra at increasing cooling power gives a reliable estimate of $\nm$, skipping the independent calibration of the optomechanical coupling, we see that  uncertainties in additional noise sources can reduce the measurement accuracy. Therefore, the measurement of the motional sidebands ratio in heterodyne spectra remains in principle a superior procedure. Indeed, it gives directly access to the real average phonon occupation number for each value of cooling power, including implicitly extra heating and noise and without the necessity of further independent measurements of system parameters. 

\begin{figure}
\begin{centering}
\includegraphics[scale=0.4]{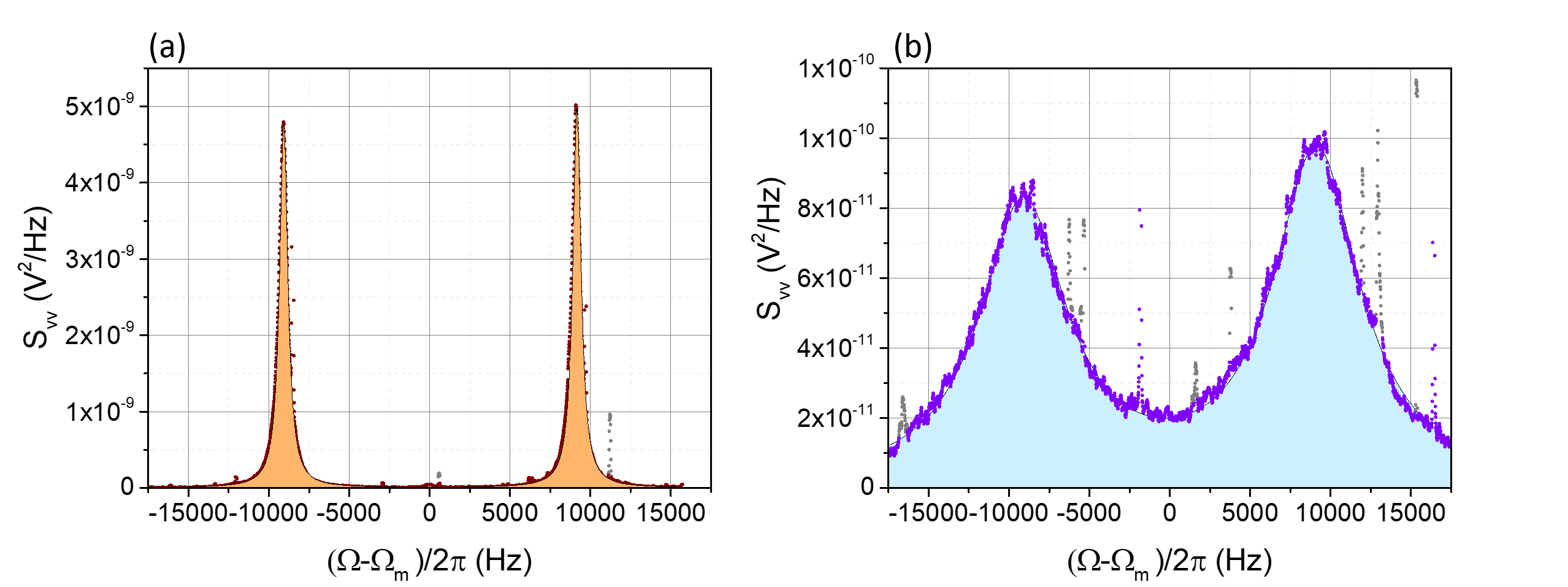}
\par\end{centering}
\caption{Observation of the Stokes (right) and anti-Stokes (left) spectral peaks of the (1,1) membrane mode for two different
values of the cooling power: (a) At low cooling power the spectral width $\Geff$ is still relatively small and the sideband asymmetry
is just $R \simeq 1.048$, yielding an inferred mean phonon occupancy of
$17.1\pm3.4$. (b) At larger cooling power, producing broader peaks, the 
asymmetry is more evident, with a measured value of $R \simeq 1.24$
and a mean phonon occupancy of $3.87\pm0.21$. Symbols show the experimental data, including the narrow peaks of the ``heavy twin'' mode and spurious electronic peaks shown in light grey. Solid lines are the fitting functions, composed of couples of Lorentzian peaks of equal width and shifted by $2 \Delta_{\mathrm{LO}}/2 \pi$, plus a linear backround that is subtracted from the displayed data for the sake of clarity. The fitted mean resonance frequency is taken as origin of the displayed horizontal axis.}
\label{hetero}
\end{figure}

Our setup
can easily switch from homodyne to heterodyne detection, by just including a frequency offset $\Delta_{\mathrm{LO}}$ in the phase locking of the local oscillator.  
Figure \ref{hetero} shows two examples of heterodyne spectra, again around the resonance frequency of the (1,1) modes, for two different values of the cooling
power. At low power (panel a) the motional sidebands are very similar, while at higher cooling power (panel b) the increased width $\Geff$, indicating a smaller occupation number, is accompanied by a visible asymmetry, with a smaller left (anti-Stokes) sideband. For a correct evaluation of $\nm$ one must consider the filtering effect of the cavity, and in particular evaluate the residual probe detuning. To this purpose, we have exploited the motional sidebands of the ``heavy twin'' mode that, being weakly coupled to the optical field, maintains a occupation number so high that a possible sideband asymmetry should be completely attributed to the cavity filtering effect. Operatively, we measure the sidebands ratio $R_{light}$ for the ``light twin'' mode and $R_{heavy}$ for the ``heavy twin'' mode, and correct the former according to $R = R_{light}/R_{heavy}$. We use this corrected $R$ to estimate $\nm$, that assumes, e.g., the value of $\nm=17.1\pm3.4$ for the spectra in panel (a) and $\nm=3.87\pm0.21$ for panel (b) (the reported uncertainty is the standard deviation in 10 measurements, performed on consecutive, 10 s long time intervals, for a total measurement time of 100 s). The latter value is obtained for our maximum cooling rate.
\begin{figure}
\begin{centering}
\includegraphics[scale=0.6]{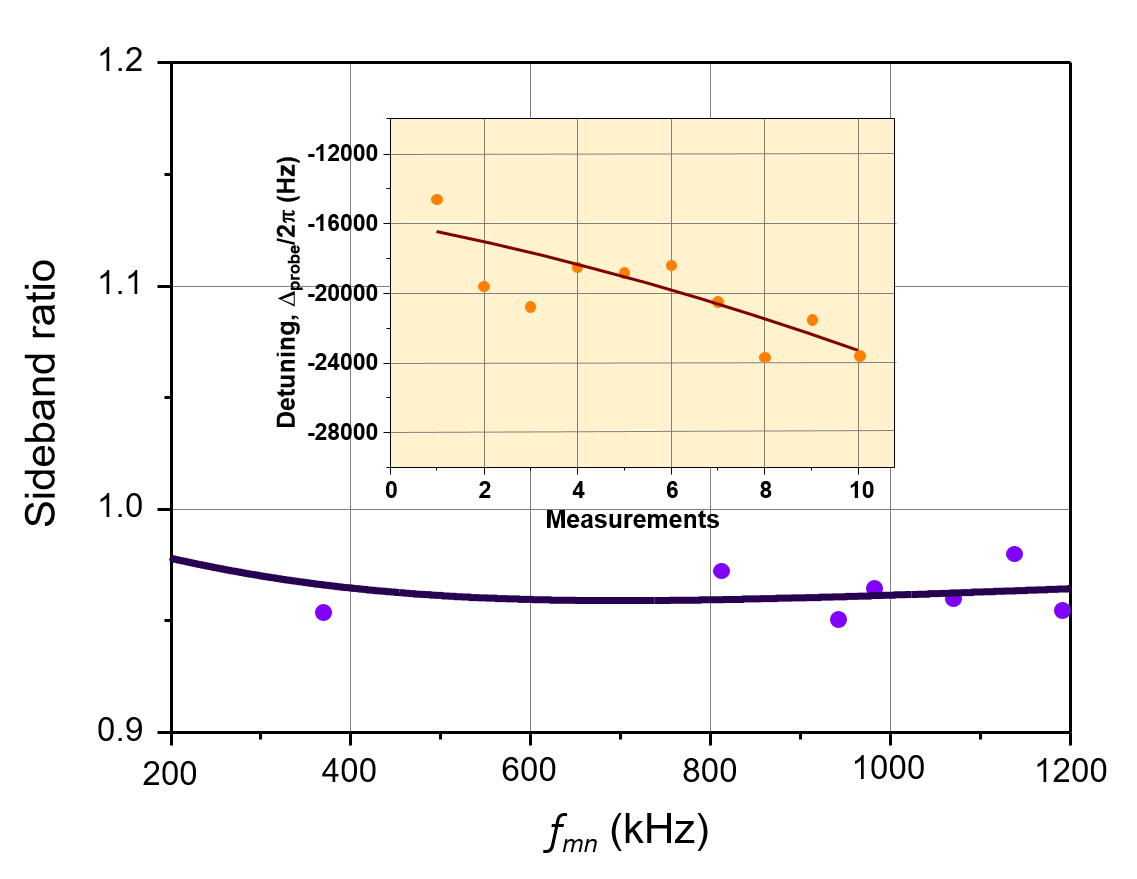}
\par\end{centering}
\caption{Method for correcting the sideband asymmetry due to the residual probe detuning. The measured sideband ratio
for several weakly coupled modes is plotted as a function of the respective resonance frequencies $\Omega_m$ (blue dots), and fitted with the function $\mathcal{L} (\Delta_{probe} - \Omega_m)/\mathcal{L} (\Delta_{probe} + \Omega_m)$ to infer the probe detuning $\Delta_{probe}$ (solid line). This procedure is repeated for several consecutive, 10 s long time intervals. The evolution of the inferred values of the detuning (shown with orange close circles in the inset)
is fitted with a first or second order polynomial function (solid line in the inset).}
\label{correction}
\end{figure}

This calibration procedure relies on the close proximity of the resonance frequencies of the two (1,1) modes, yielding the same cavity filtering effect. However, one can also evaluate the sideband ratio for several weakly coupled modes, deduce the probe detuning $\Delta_{probe}$ by fitting the results with the function $\mathcal{L} (\Delta_{probe} - \Omega_m)/\mathcal{L} (\Delta_{probe} + \Omega_m)$ vs $\Omega_m$, and finally use the same function with the inferred $\Delta_{probe}$ and $\Omega_m = 2 \pi \times 370\,$kHz to correct $R_{light}$. An example of such fit is shown in Fig. \ref{correction}. This procedure 
also allows to monitor the stability of the detuning during the measurement, as shown in the inset of Fig. \ref{correction}. We have found typically a detuning below $\mid \Delta_{probe}\mid /2\pi < 30$ kHz (corresponding to $0.02\kappa$) and variations during a complete measurement three times smaller. The consequent corrections to $R_{light}$ 
arrive to nearly $10 \pct$. A preliminary evaluation of the sideband ratio for the weakly coupled modes is indeed a good method to adjust the probe detuning at the beginning of the measurement. The corrections to $R_{light}$ obtained with 
this procedure are in good agreement with the method using directly the ``heavy twin'' mode.      

\begin{figure}
\begin{centering}
\includegraphics[scale=0.6]{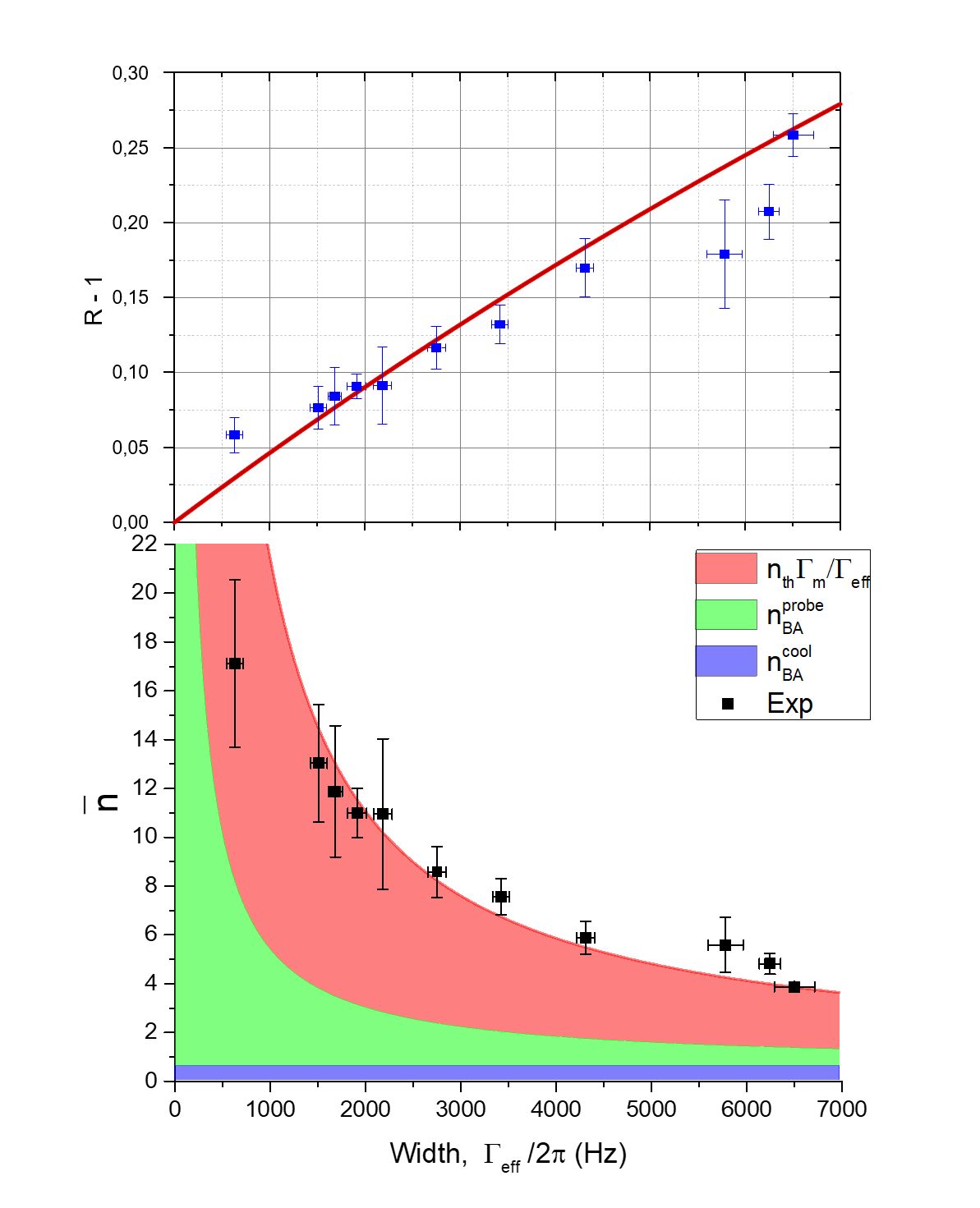}
\par\end{centering}
\caption{Close symbols report the occupation number $\nm$ calculated from the corrected values $R$ of the sideband ratio for the ``light twin'' mode, according to $\nm = 1/(R-1)$. Error bars reflect one standard deviation in the measurements performed on 10 consecutive, 10 s long time intervals. The red solid curve represents the occupation number $\nm$ calculated according to Eq. (\ref{ntot}) using independently measured parameters. Red, green and blue areas represent respectively the contributions of the thermal noise, the probe beam back-action, and the cooling beam back-action.}
\label{fignm}
\end{figure}
The occupation number $\nm$ calculated from the corrected sideband ratio is shown in Figure \ref{fignm} as a function of $\Geff$,
obtained at increasing values of cooling power. Filled solid curves reflect the expected $\nm$ and its different contributions, calculated according to Eqs. (\ref{nbac}-\ref{ntot}). In particular, $n_{BA}^{cool}\sim0.58$, showing that for the (1,1) modes we are here in weakly resolved sidebands regime and back-action cooling can in principle bring these modes to an occupation number below unity (close to $n_{BA}$ in the weak coupling regime \cite{Marquardt2007}. With respect to the analysis of the results extracted from the homodyne spectra, described in Fig. {\ref{homo}b, here the theoretical curves have no free fitting parameters: all the contributions to $\nm$ are calculated on the basis of independent measurements. The agreement with the experimental data is excellent, considering the experimental statistical uncertainty, suggesting the absence of non-modeled extra noise. Each single data point can thus be exploited to extract the occupation number, using as experimental error its statistical uncertainty that, differently from the case of $\area \Geff$, is now compatible with the data scattering. On the other hand, the overall set of data could be used evaluate $\Temp$, leaving $\nth$ as free parameter in the expression (\ref{ntot}). In this case, the extracted value is $6.7 \pm 0.6$ K, compatible with the 7.2 K measured by the sensor.

\section{Conclusions}

We compare two indicators of the oscillator occupation number, namely the peak area$\times$width product of the displacement spectrum, acquired in a homodyne setup, and the motional sidebands asymmetry, measured by heterodyne detection. Neither case requires additional calibrations, even if an additional absolute calibration of the homodyne spectrum in terms of frequency fluctuations allows to additionally infer the single-photon optomechanical coupling strength $g_0$. Both indicators are particularly sensitive at low occupation numbers (i.e., in the quantum regime). In our case the two kinds of estimate are in agreement, showing that a minimal occupation number of 3.9 is achieved in our experiment. However, the latter indicator is superior because it is less sensitive to additional technical noise, and it gives a result with a single measurement while the former procedure requires a set of measurements as a function of, e.g., the cooling power. 

To reliably exploit the latter indicator one should keep in mind that a crucial requirement for an accurate measurement of the sidebands ratio is the control of the probe detuning. We show a method to perform it, based on the observation of the spectral features of weakly coupled mechanical mode. The calibration of the detuning is thus performed using phase signals generated inside the optomechanical cavity. This method is more trustworthy than those exploiting calibration tones on the probe field since simultaneous phase and amplitude modulation, that commonly occurs in real setups, easily generates asymmetric sidebands that spoils accurate measurements of the cavity detuning.

A widespread use of reliable quantum optomechanical indicators, toward which this work is contributing, is expected to play in important role in the development of quantum technologies \cite{Acin2018}.        

\ack

Research performed within the Project QuaSeRT funded by the QuantERA ERA-NET
Cofund in Quantum Technologies implemented within the European Union's
Horizon 2020 Programme. The research has been partially supported by INFN
(HUMOR project). AP has received funding from the European Unions Horizon 2020 research and
innovation programme under the Marie Sklodowska-Curie
Grant Agreement No. 749709.

\end{document}